# 7. Einstein en la Argentina: el impacto científico de su visita

Alejandro Gangui y Eduardo L. Ortiz

## 1. Introducción

En paralelo con los viajeros argentinos que se trasladaban a Europa, para estudiar en sus centros de cultura, nuestra historia registra la presencia de científicos visitantes a lo largo de un período muy dilatado, que se remonta a los tiempos de la Colonia cuando estos visitantes fueron en su gran mayoría naturalistas o cosmógrafos. Con la Independencia no se interrumpió esta cadena particular de visitantes sino que, por el contrario, fue fuertemente estimulada. Una de las primeras acciones de Rivadavia en Europa fue invitar a científicos destacados para que enseñaran en nuestra capital. Así llegaron a nuestras playas primero un matemático de gran renombre en Europa, José María Lanz, y luego un físico igualmente eminente, Ottaviano-Fabrizio Mossotti. Más tarde llegaron también a la Argentina otras personalidades científicas, entre quienes se destacan Aimé Bonpland y Charles Darwin.

Entretanto, por los avances en la seguridad y rapidez de la navegación a vela, particularmente con la difusión de los *clippers* en la navegación hacia el sur, comenzaron a hacerse posibles visitas de personalidades europeas a las Américas por espacios considerablemente más breves que en el pasado y que ahora se medían en términos de meses, y aun de semanas, no ya de años. En el campo de la ciencia y la tecnología, por ejemplo, una visita muy interesante fue la que realizó a Buenos Aires el ingeniero inglés John Bateman en 1870-71. Esa visita de asesoramiento, que se extendió por menos de dos meses, dejó resultados decisivos para el diseño del puerto, del sistema de suministro de aguas y de la salubridad de esta ciudad. La visita fue el heraldo de un nuevo tipo de encuentros, breves pero capaces de producir un impacto considerable, que comenzaron a realizarse sobre todo entre fines del siglo XIX y comienzos del XX. Este fue el momento en que las realizaciones de la ciencia y la tecnología habían conseguido crear en la Argentina una audiencia suficientemente amplia y cuando el triunfo del vapor sobre la vela inauguró servicios más regulares, confiables y económicos entre Europa y la América del Sur.

Así fue que con motivo de las celebraciones del Centenario de 1910, entre las diversas personalidades que visitaron Buenos Aires llegaron dos eminentes científicos. Uno de ellos, que acompañaba a la representación española, fue Leonardo Torres Quevedo, uno de los artífices más destacados del mundo moderno de la automatización y de la computación, que explicó el funcionamiento de una de sus máquinas electromecánicas de cálculo automático. El otro, que llegó junto a la representación italiana, fue el matemático y físico teórico Vito Volterra, uno de los primeros relativistas. En una conferencia histórica, Volterra presentó las modernas ideas que vinculaban el espacio, el tiempo y la materia y mencionó el nombre del joven Albert Einstein (Volterra, 1910). El vapor y el Centenario marcaron también el comienzo de una serie de visitas breves de intelectuales españoles –muchos de ellos entrenados en Alemania– auspiciadas por la benemérita Institución Cultural Española de Buenos Aires (Ortiz, 1988). Entre ellos se encontraban tanto el filósofo José Ortega y Gasset,[1] como el matemático Julio Rey Pastor y el físico Blas Cabrera, quien dictó una serie de conferencias sobre la relatividad en su visita de 1920 (Ortiz y Rubinstein, 2009).

Este tipo de conferencias sobre la relatividad, así como diversos trabajos producidos localmente sobre la materia, venían preparando de alguna manera el terreno para la visita de Albert Einstein (1879-1955) en 1925.[2] Por empezar, estudios sobre el espaciotiempo y la radiación, muy anteriores a los trabajos de Einstein, comenzaron a aparecer en Europa y luego en nuestro país, hacia fines del siglo XIX, en trabajos del matemático Valentín Balbín y del ingeniero y físico teórico Jorge Duclout, entre otros autores. En la primera década del siglo XX se publicaron también notas sobre la materia y la radiación y sobre la nueva teoría de la relatividad (restringida) de Einstein, que parecía ofrecer una visión unificada de la materia y la energía e introducía concepciones nuevas sobre la estructura del espacio físico. Esos estudios producidos en el ambiente local tenían un carácter informativo serio y se difundieron en los medios de prestigio científico del momento: revistas universitarias, revistas de ingeniería, y en los *Anales de la Sociedad Científica Argentina* (Galles, 1982).

Por otro lado, se destaca el trabajo del matemático italiano Ugo Broggi, aparecido en 1909, sobre la materia, la radiación y el tiempo. Entre 1910 y 1915 el físico-matemático francés Camilo Meyer dictó una serie de cursos libres sobre física-matemática en la Universidad de Buenos Aires y en el último de ellos introdujo por primera vez en América Latina una exposición detallada de la teoría cuántica (Meyer, 1915). En ese curso hizo referencias explícitas a trabajos de Einstein pero, por lo específico de su tema, no a aquellos que se relacionaban directamente con la teoría de la relatividad. Por su parte, el prestigioso astrónomo estadounidense Charles Dillon Perrine, director del Observatorio de Córdoba, respondiendo a un pedido de colegas suyos en Alemania asociados con Einstein, inició en 1912 una serie de observaciones durante eclipses de Sol tendientes a detectar una posible deflexión de la luz. Esas observaciones comenzaron en Brasil en 1912 y continuaron luego en Rusia y Venezuela, pero debido a condiciones meteorológicas desfavorables, Perrine y sus colaboradores no pudieron ofrecer una respuesta concluyente (Gangui y Ortiz, 2009). Las observaciones buscadas por los astrónomos para verificar la teoría general de la relatividad solo llegarían en el año 1919 pero, lamentablemente, la Argentina no se hizo presente en esa ocasión tan favorable. Cabe destacar, como otro aporte a la difusión de la teoría de la relatividad, el trabajo del astrónomo jesuita español José Ubach (profesor de ciencias en el Colegio del Salvador de Buenos Aires, donde mantenía un observatorio) sobre los resultados de la expedición astronómica de 1919.

---

[1] Ver el capítulo de Maximiliano Fuentes Codera, "José Ortega y Gasset y Eugenio d'Ors: las primeras visitas a la Argentina y sus proyecciones", y el de Darío Roldán, "La segunda visita de José Ortega y Gasset a la Argentina", en este mismo libro.

[2] Más detalles sobre esta visita se encuentran en Ortiz, 1995; Ortiz y Otero, 2001; Gangui y Ortiz, 2005; 2008; 2009; 2011; Asúa y Hurtado de Mendoza, 2006 y Gangui, 2007.

También en 1912, el físico Jacobo Laub comenzó a publicar una serie de trabajos sobre la teoría de la relatividad en *Anales de la Sociedad Científica*. Las suyas fueron las contribuciones de un científico que conocía las ideas de Einstein y su teoría en detalle (Gangui y Ortiz, 2011), dado que antes de ser contratado en 1910 por la Universidad Nacional de La Plata, había tenido un contacto personal con Einstein (Seelig, 1954) y había colaborado con él en trabajos de investigación sobre la teoría de la relatividad (Pyenson, 1985). A medida que la década avanzaba y el interés público por la relatividad crecía, sus notas comenzaron a ser publicadas en medios de difusión más ámplia como la *Revista de Filosofía* que dirigía José Ingenieros.

Por último, unos años más tarde apareció un interesante estudio sobre la teoría de la relatividad escrito por dos jóvenes físicos argentinos, José B. Collo y Teófilo Isnardi que habían hecho estudios avanzados de física en Alemania. Por la calidad de su factura, este trabajo –que incluía una tercera parte escrita por el astrónomo Félix Aguilar– resulta paradigmático, dentro de las ciencias físicas, de un nuevo e importante giro intelectual hacia las ciencias teóricas que se operaba en la Argentina de esos años (Gangui y Ortiz, 2009; 2011).

Pese a la difusión que había alcanzado la teoría de la relatividad tanto en medios especializados como en otros que lo eran menos, la complejidad de las ideas de Einstein hizo que resultara compleja la difusión de su mensaje en la prensa. En realidad, si bien los científicos que visitaban el país dejaban su marca en la prensa –sobre todo porque daban seguridades de que el progreso era realmente indefinido–, al mismo tiempo, comunicarse con ellos o entender su mensaje era a veces difícil. Para aquellos que eran ya suficientemente famosos existían, en las oficinas de los grandes diarios locales, legajos de recortes de la prensa extranjera que facilitaban la recepción de sus ideas y permitían repetir sobre ellos, con cierto margen de seguridad, lo que se había dicho sobre sus visitas a alguna de las grandes capitales de Europa. Pero el caso de Einstein resultó en un punto desconcertante: no solo por la complejidad de su discurso, sino también por su apariencia, que no encajaba con la imagen del sabio visitante firmemente asentada en la mente de sus interlocutores. Su apertura y su tendencia a ironizar, sobre todo en sus primeras entrevistas con la prensa, desorientó primero a los periodistas y más tarde también a algunos miembros de su audiencia intelectual. Einstein, entonces, obligó a revisar tanto la imagen del sabio como la del espaciotiempo.

El arribo de Albert Einstein a la Argentina conmocionó por igual a los círculos científicos y a la opinión pública. Este trabajo se propone discutir esta visita en tres momentos diferentes. Primero, hace un repaso de los condicionantes que permitieron planear el viaje y de los actores institucionales que lo facilitaron. A continuación, discute algunas actividades sociales y científicas que desarrolló el visitante durante el largo mes que permaneció en el país y los contactos que mantuvo con referentes locales. Finalmente, considera las repercusiones que la estadía de Einstein tuvo en el ambiente científico local y las referencias posteriores a su visita que aparecieron en medios locales e internacionales.

### 2. La figura pública de Einstein en la Argentina

El año 1919 es recordado como un momento de inflexión en la figura pública de Einstein. El 29 de mayo de ese año se produjo un eclipse total de Sol que permitió poner a prueba una de las predicciones más importantes de la teoría de la relatividad general: la deflexión de la luz en un campo gravitatorio. Luego de propuesta la teoría a fines de 1915 y de varios intentos frustrados de verificación astronómica, en 1917 el astrónomo Real de Inglaterra, Frank Watson Dyson, sugirió que se organizara una nueva expedición para estudiar el eclipse total de 1919 y que se intentara verificar la novedosa teoría propuesta por Einstein. Con ese fin se enviaron dos expediciones: una a Sobral (estado de Ceará, en el norte de Brasil), liderada por los astrónomos Andrew Crommelin y Charles Davidson; la otra a la isla Príncipe, en África Occidental, frente a la costa de la Guinea Española, encabezada por otros dos astrónomos, Edwin Cottingham y Arthur S. Eddington (una figura de gran prestigio que se había ocupado en detalle sobre los estudios de Einstein). El resultado de las delicadas observaciones astronómicas fue el esperado por los expedicionarios: las estrellas estudiadas en la vecindad del disco solar tenían en efecto sus posiciones desplazadas del centro de dicho disco en aproximadamente la cantidad (ínfima) predicha por la teoría de la relatividad, según la cual el espacio, y por lo tanto también las trayectorias de los haces de luz de las estrellas, se "curvan" debido a la presencia del Sol.

En Inglaterra, el anuncio de los resultados se hizo en una ceremonia especial: se convocó a un congreso conjunto de las prestigiosas Royal Society y Royal Astronomical Society en Londres para el 6 de noviembre de 1919. Resulta clara la relevancia de la reunión: poco después del fin de la Primera Guerra Mundial, un grupo de astrónomos ingleses liderado por el pacifista y *quaker* Eddington presentaba pruebas fehacientes de que una nueva teoría, surgida de la mente de otro

pacifista –pero alemán– debía reemplazar a la vieja mecánica de Newton vigente desde hacía más de dos siglos.

A partir del "eclipse de Einstein" comenzaron a aparecer con frecuencia noticias y artículos sobre diversos aspectos de la teoría de la relatividad y sobre las discusiones científicas en Europa sobre temas relacionados. La Argentina no se mantuvo ajena a ese interés y es indudable que Leopoldo Lugones fue uno de los principales responsables de la presencia de estos temas en los medios locales. El interés de Lugones por la ciencia, en particular por la teoría de la relatividad, está bien documentado y, por otra parte, el escritor tenía una influencia intelectual destacada sobre los editores de varios de los periódicos de gran circulación en el país. Aunque no se trataba de un experto, era un intelectual respetado y su juicio sobre casi cualquier tema tenía un peso considerable. En 1920, fue invitado a dictar una conferencia sobre las implicancias de la teoría de la relatividad por el Centro de Estudiantes de Ingeniería de la Facultad de Ciencias Exactas, Físicas y Naturales –conocida entonces como "Facultad de Ingeniería"–, una institución estudiantil que jugó un papel destacado en el panorama cultural de la primera mitad del siglo XX. En base a esa conferencia, al año siguiente Lugones publicó *El tamaño del espacio* (Lugones, 1921). Aunque no estaba libre de interpretaciones erróneas, de citas algo extemporáneas o de párrafos oscuros y aun esotéricos, el libro sirvió para posicionar favorablemente a la teoría de la relatividad frente a la audiencia que se interesaba por los temas científicos en el Buenos Aires de la época. El solo hecho de que un intelectual del calibre de Lugones se interesase por esta teoría era en principio mucho más importante para su recepción que la precisión en la descripción de la física (Ortiz, 1995). En el diario *La Nación* del 23 de junio de 1921, se señalaba que Einstein, dirigiéndose a un reportero español, agradeció a Lugones por la gentil referencia a su persona. Ese año y los siguientes, los principales diarios de Buenos Aires incluyeron de forma habitual noticias acerca de los múltiples tributos que el sabio alemán recibía en diferentes países del mundo, particularmente en Francia, un norte en la Argentina para la adjudicación de prestigio.

Un mes más tarde que Lugones, y también invitado por el Centro de Estudiantes de Ingeniería, Duclout leyó una ponencia sobre materia-energía y relatividad que luego fue publicada (Duclout, 1920). El autor, que se había afincado en el país en la década de 1880, tenía una cultura física y matemática sólida, se había graduado en el prestigioso Politécnico de Zürich –al igual Einstein– y tuvo un rol importante en la materialización de la visita del sabio alemán. Por su parte, el matemático puro Julio Rey Pastor (Ortiz, 2011a), de origen español pero entrenado en Alemania y, en ese momento, residente ya en nuestro país, contribuyó con trabajos relacionados con la relatividad, principalmente en su relación con la geometría. Más cerca de la visita de Einstein, publicó notas divulgativas sobre tópicos relacionados con la relatividad en diarios locales, particularmente en *La Nación*. Rey Pastor había conocido a Einstein en Berlín y jugó un papel importante en su invitación a visitar España en 1923.

También los estudiantes universitarios manifestaron interés por los temas que habían hecho célebre a Einstein. Nuevamente fueron las autoridades del Centro de Estudiantes de Ingeniería las que, aparentemente por sugerencia de Rey Pastor, escribieron a Einstein el 5 de abril de 1922 pidiéndole permiso para traducir y publicar el trabajo técnico "Die Grundlage der allgemeinen Relativitätstheorie" –originalmente aparecido en la revista alemana *Annalen der Physik* en 1916– y el argumento esgrimido para tal empresa era el atractivo que suscitaban esos tópicos en los círculos universitarios locales. Einstein respondió a ese pedido el 31 de mayo de ese mismo año aceptando la propuesta y sugiriendo que la publicación fuera en forma de libro y que le reservaran un 20% de las ganancias en calidad de derechos de autor.

### 3. Repercusiones locales frente a las amenazas a Einstein

Para el año 1922 la situación social en Alemania había comenzado a degradarse. Los periódicos de la Argentina reiteraban sus noticias sobre la inestabilidad política de este país y, a fines de junio de ese año, anunciaron el asesinato de Walther Rathenau, ministro de Relaciones Exteriores de la República de Weimar. *La Nación* también trajo a estas tierras la noticia de que la vida de Einstein corría peligro: hacia fines de 1923, debió abandonar su país y viajar temporariamente a Holanda por las amenazas que había recibido.

Ante esta situación, el 9 de agosto de 1922 Lugones publicó en *La Nación* un artículo donde sugería que el sabio fuese invitado a la Argentina y que sus admiradores locales se encargasen de contribuir con fondos para cubrir los gastos y crear una cátedra universitaria independiente especial para él. Poco tiempo después de esta nota, dos agrupaciones estudiantiles de Buenos Aires, el Centro de Estudiantes de Ingeniería y el del Instituto Nacional del Profesorado Secundario, anunciaron públicamente su apoyo a la idea de Lugones. También lo habrían hecho los científicos Rey Pastor, consejero de la primera asociación y Laub, en ese entonces jefe del Departamento de Física del Instituto Nacional del Profesorado Secundario. Sin duda alguna, Duclout, como amigo personal de Lugones y Rey Pastor, también fue parte de este movimiento de opinión.

Al mismo tiempo, el 22 de agosto de ese año Duclout presentó ante el Consejo Directivo de la Facultad de Ciencias Exactas, Físicas y Naturales la propuesta de otorgar a Einstein el título de doctor *honoris causa* en ciencias físico-matemáticas por sus trabajos sobre la teoría de la relatividad. Avalada por unanimidad, en solo unos pocos días su propuesta fue aceptada también por el Consejo Superior de la Universidad de Buenos Aires. El 30 de octubre de ese mismo año, secundado por otros académicos, fue nuevamente Duclout quien presentó ante el Consejo de la UBA una propuesta formal para que Einstein fuese invitado a dictar una serie de conferencias sobre la teoría de la relatividad. La ocasión no podía haber sido mejor elegida, pues menos de dos semanas más tarde los matutinos porteños anunciaban que se había otorgado a Einstein el premio Nobel de física correspondiente al año 1921, que había quedado vacante del año anterior. Cabe recordar que el científico alemán no recibió ese premio por sus trabajos sobre la relatividad, que no era universalmente aceptada aun después de las pruebas astronómicas, sino por su explicación del llamado efecto fotoeléctrico. Esa era una época de viajes frecuentes para Einstein: había sido invitado a visitar Inglaterra y Estados Unidos en 1921, Francia y Japón en 1922 y España y Palestina el año siguiente. La prensa argentina seguía en ese entonces, con detalle y con un marcado interés, todos sus movimientos.

Un año más tarde, el 21 de diciembre de 1923, el Consejo Superior de la Universidad de Buenos Aires se reunió en sesión extraordinaria. Entre sus temas de discusión figuraba, nuevamente, la visita de Einstein y se tomó en consideración una nota de la Asociación Hebraica (luego, Sociedad Hebraica Argentina), donde señalaba que se había contactado con el físico para que llevara a cabo un *tour* de conferencias por la Argentina.[3] Finalmente, el Consejo Superior aprobó un presupuesto para lograr la visita del físico, al que se sumaron los aportes de la Asociación Hebraica y de la Institución Cultural Argentino-Germana. A partir de ese momento, el Consejo Superior de la Universidad recibió noticias frecuentes del estado de las negociaciones con el futuro visitante. En mayo de 1924, fue informado de la correspondencia que Einstein mantuvo, por un lado, con el representante del Ministerio de Relaciones Exteriores de la Argentina en Berlín y, por otro, con la Asociación Hebraica. La fecha fue fijada para fines de marzo del año siguiente, es decir, hacia el comienzo del año académico de 1925.

Es interesante señalar que, para explicar la teoría de la relatividad, la pujante Argentina de esos años buscó nada menos que la palabra del creador de esa teoría. Unos pocos años más tarde, llegó a Buenos Aires otro relativista eminente, Paul Langevin que, con el auspicio del Institut Français del'Université de Paris à Buenos Aires visitó el país con un costo sumamente reducido para la Universidad; además, rechazó los honorarios especiales que esta le ofreció por sus conferencias.

### 4. Einstein en la Argentina

A los 46 años de edad, Albert Einstein emprendió un largo viaje en barco que lo llevó a Sudamérica y en el marco del cual permaneció en Buenos Aires durante todo un mes y realizó breves visitas a otras ciudades argentinas. Esta experiencia, así como muchas otras de su abultada carrera científica y humanista, quedaron plasmadas en sus diarios íntimos, donde Einstein dio cuenta de muchos detalles que no salieron reportados en los diarios de la época. El diario correspondiente a su viaje a Sudamérica permite reseñar algunos aspectos de su paso por la Argentina.

---

[3] Sobre la relación de Einstein con la comunidad judía, ver (Ortiz, 1985) y el capítulo de Alejandro Dujovne, "Einstein y la comunidad judía argentina", en este mismo libro.

Como es de imaginar, durante su visita un científico de la talla de Einstein no podía pasar desapercibido para la sociedad local. Por el contrario, Einstein fue abrumado con interminables entrevistas, banquetes y honores y no tuvo demasiado tiempo para trabajar o para descansar. Llegó al puerto de Buenos Aires al alba del día 25 de marzo de 1925 pero su periplo había comenzado casi tres semanas antes, cuando abandonó el puerto de Hamburgo a bordo del navío alemán *Cap Polonio*, que traía a Einstein en su lujosa primera clase y a inmigrantes, en su mayor parte judíos, en una congestionada y algo menos acogedora tercera clase. Contrariamente a lo que se ha afirmado en varias publicaciones, y aunque disponía de dos pasajes, Einstein no viajó acompañado de su esposa. Luego de una escala breve en Rio de Janeiro, realizó el trayecto de Montevideo a Buenos Aires acompañado por miembros de diversas comitivas de bienvenida que habían ido a esperarlo a la otra orilla del Plata. Sus integrantes eran científicos, personalidades académicas y miembros de la comunidad judía de Argentina y entre ellos se encontraba el Secretario de la Universidad de Buenos Aires, Mauricio Nirenstein, miembro conspicuo, a la vez, de la Asociación Hebraica.

A su llegada al puerto de Buenos Aires, Einstein recibió una primera muestra del impacto público de su visita: una masa de periodistas, cámaras y filmadoras lo aguardaban en la dársena de desembarco. Apresurados por esquivarlos, los acompañantes de Einstein intentaron evadirse con él en un automóvil particular, pero fueron firmemente detenidos por los periodistas, que usaron el propio equipaje de los viajeros para bloquear el camino y no los dejaron partir hasta haber fotografiado y filmado al visitante en el interior del vehículo. Einstein escribía en su diario íntimo: "A las 8:30 estábamos en tierra firme. Nirenstein me prestó ayuda" y agregaba que se sentía asfixiado y "medio muerto"[4] luego de tanto viaje y ajetreos entre la multitud. Finalmente, llegó a la residencia de Bruno Wassermann, un rico comerciante papelero de origen judío-alemán, situada en la parte más elegante de Belgrano, donde se alojó durante su estadía en Buenos Aires. Allí, Einstein pudo tomarse un brevísimo descanso y escribió en su diario: "Tranquilidad por fin, estoy totalmente deshecho".[5] Una vez que Einstein llegó a Buenos Aires, Nirenstein compartió la secretaría universitaria con la función de secretario del ilustre visitante, o quizás más bien de severo preceptor (Gangui y Ortiz, 2008).

---

[4] A. Einstein Archive, Princeton University, Doc. 29-132, p. 14.
[5] A. Einstein Archive, Princeton University, Doc. 29-132, p. 14.

Las inevitables visitas protocolares no se hicieron esperar. Dejando de lado a los periodistas, la lista de los intelectuales que lo visitaron fue encabezada, como era de esperarse, por Leopoldo Lugones, uno de los pocos argentinos, quizás el único intelectual del país, a quien Einstein había conocido antes de su viaje. Ambos habían compartido, junto con Marie Curie, Henri Bergson y otras figuras eminentes, la mesa de discusiones de uno de los foros intelectuales más importantes de esa época, el Comité Internacional de Cooperación Intelectual de la Liga de las Naciones –un embrión de UNESCO–, donde Lugones representaba a la Argentina.

Siguió luego una larga lista de personalidades con las que se entrevistó: el embajador alemán, las autoridades de la Universidad de Buenos Aires, algunos científicos de renombre, representantes de la Universidad de La Plata y de organizaciones judías y no judías (como la mencionada Asociación Cultural Argentino-Germana) que habían colaborado en la financiación de su circuito en el país y, con éxito variable, habían intentado extenderlo a otros lugares de Sudamérica. El diario *La Prensa* describió a Einstein utilizando un estereotipo desarrollado ya por el periodismo de otros países: "bondadoso, afable y simpático", "verdadero prototipo de la ciencia", famoso por ser el "autor de una teoría científica que ha llamado la atención del mundo". A estos atributos se agregarían luego su humildad, la pobreza de su vestimenta y el rol que en su presencia física jugaba su impresionante y desplegada cabellera. En declaraciones al matutino, Einstein agradeció "la elogiosa crónica que sobre mi persona publicó ayer el gran diario argentino" y manifestó que era "enemigo del exhibicionismo" y que "me consideraría muy satisfecho si no se me abrumara tanto con el sinnúmero de entrevistas que se me solicitan"[6]. Pero la maratón recién comenzaba. El jueves 26 de marzo, por la mañana, Einstein recibió a periodistas y fotógrafos en la residencia de los Wassermann. Al mediodía, lo llevaron a conocer dos caras diferentes de la ciudad: los cuidados bosques de Palermo y el Mercado de Abasto Proveedor, que Einstein cruzó a pie con su comitiva. Esta última visita evocaba la imagen del tango argentino que comenzaba ya a atraer la atención de Europa. Por la tarde, asistió a una recepción formal en el salón de actos del Colegio Nacional de Buenos Aires, donde pronunció su conferencia introductoria en francés, ante un auditorio ruidoso, amplio y numeroso donde no estuvo ausente el público femenino. Su discurso fue breve y se limitó a consideraciones de carácter general; en realidad, Einstein había preparado un texto especial que no fue el que leyó y que permaneció durante mucho tiempo inédito (Einstein, 2008). Luego de entrevistas periodísticas muy abiertas y de un artículo para la prensa local sobre la Pan-Europa –una aspiración de los pacifistas de la postguerra concretada muchas décadas más tarde con la Unión Europea–, los asesores locales de Einstein parecen haberle aconsejado limitar sus intervenciones a la física teórica (Gangui y Ortiz, 2008).

---

[6] "El profesor Einstein, en tierra argentina, formuló interesantes declaraciones para 'La Prensa'". *La Prensa*, Buenos Aires, 26.03.1925.

El viernes por la noche, fue invitado a una recepción en la residencia de un miembro influyente y adinerado de la comunidad judía-alemana local, Alfredo Hirsch, instalado en el país desde hacía tres décadas como uno de los más fuertes comerciantes en el ramo de la exportación de cereales. Allí, Einstein se encontró con personas que habían apoyado financieramente su viaje a Argentina. En su diario íntimo, hizo referencia, entre elogiosa e irónica, a las riquezas que adornaban a esa residencia, sin olvidar una justificada referencia a la belleza de la esposa de Hirsch.

Su primera lección estrictamente científica en Buenos Aires (de las varias conferencias dictadas en diferentes centros académicos) se realizó el sábado 28 de marzo. Ese día la sala estaba colmada, según escribía Einstein en su diario, donde agregaba que "La juventud es siempre agradable y se interesa por las cosas"[7]. A esta conferencia, además de las autoridades universitarias del más alto rango, asistieron dos ministros del gabinete, los titulares de las carteras de Educación y de Relaciones Exteriores y varios embajadores extranjeros.

Al día siguiente, el científico logró "permanecer solo en [su] habitación durante la mañana" y disfrutar de una jornada de "feliz tranquilidad"[8] y, por la tarde, realizó una caminata en compañía de Wassermann. Pero, llegado el lunes, el protocolo continuó con paso firme: al mediodía visitó los modernísimos talleres del diario *La Prensa* y luego dictó su segunda conferencia científica sobre relatividad, con abundante tiempo para discusiones. En esta ocasión, varios científicos locales intentaron mostrarse capaces de hacer preguntas, no siempre con éxito. Según la opinión de algunos de los asistentes, Einstein improvisaba sus clases, cuya dificultad era sumamente variable (Ortiz y Otero, 2001). El martes 31, por la mañana, visitó la redacción del diario judío *Das Volk*, e hizo observaciones irónicas acerca de sus "paisanos"; visitó también escuelas, hospitales y orfanatos sostenidos por la comunidad judía; y por la tarde viajó al Tigre invitado por unos amigos suyos, suizo-alemanes.

El miércoles 1 de abril realizó, acompañado por la señora Wassermann, un vuelo corto sobre la ciudad de Buenos Aires a bordo de un *Junker* de la marina alemana que había llegado a la ciudad en vuelo de cortesía. Era su primer vuelo en avión y Einstein comentó luego cuánto lo impresionó esa experiencia, "particularmente el despegue"[9] de la aeronave. Por la tarde fue recibido por el presidente Marcelo T. de Alvear y por algunos ministros. Visitó luego el Museo

---
[7] A. Einstein Archive, Princeton University, Doc. 29-132, p. 17.
[8] A. Einstein Archive, Princeton University, Doc. 29-132, p. 18.
[9] A. Einstein Archive, Princeton University, Doc. 29-132, p. 19.

Etnológico y, poco después, dictó su tercera conferencia sobre la teoría de la relatividad. La jornada terminó con un paseo a pie por la tradicional calle Florida, del brazo de Lugones y seguido de estudiantes y curiosos. La caminata concluyó en casa del escritor, donde cenaron con su esposa. Fue la única oportunidad de Einstein de compartir una cena íntima en un hogar argentino.

El próximo día, viajó en tren a la ciudad de La Plata, donde había sido invitado a inaugurar el año académico de 1925; asistió también a una reunión científica en su honor, donde participaron el físico alemán Ricardo Gans, director del Instituto de Física de esa ciudad, y algunos de sus alumnos, varios de los cuales habían recibido entrenamiento de posgrado en física en universidades alemanas. Ese día, Einstein logró también hacerse de tiempo para visitar el "muy interesante Museo de Historia Natural"[10].

El viernes 3 de abril, fue invitado por las autoridades de la Universidad de buenos Aires a almorzar en el Jockey Club y dictó otra conferencia más de su ciclo para la Universidad. El día posterior al de cada presentación, versiones taquigráficas de sus exposiciones aparecían publicadas en los principales diarios. El sábado 4 de abril, le tocó el turno a la Facultad de Filosofía y Letras de la Universidad de Buenos Aires, donde repitió la función que dos días antes había cumplido en La Plata: la inauguración oficial de los cursos. Sin embargo, esta vez, a instancias del decano Coriolano Alberini, Einstein tomó parte activa ofreciendo un coloquio breve sobre "Las consecuencias de la teoría de la relatividad en los conceptos de espacio y tiempo". Alberini se encontraba entonces en guerra abierta con los positivistas locales y, en particular, con Ingenieros; en un artículo que publicó el 12 de abril en *La Nación* interpretó el discurso de Einstein como un auto de fe anti-positivista. Esta conferencia fue una de las pocas intervenciones de Einstein fuera del terreno estricto de la física teórica y, aun así, el discurso hacía solo una breve incursión en el terreno de la filosofía (Ortiz, 2011b; Gangui y Ortiz, 2011).

El domingo 5, con los Wassermann, Einstein viajó en automóvil a la residencia de vacaciones de la familia en Llavallol. El lunes 6 de abril continuó su contacto con los pocos científicos locales ocupados en temas de investigación original. En compañía del joven fisiólogo Bernardo A. Houssay (que 22 años más tarde sería galardonado con el Premio Nobel de Medicina), visitó el laboratorio de Eugenio Pablo Fortin, a quien Einstein describía como un "oftalmólogo y bolsista"[11] de origen francés que estaba haciendo interesantes experimentos sobre la percepción de sensaciones luminosas. Por la tarde dictó una nueva conferencia –la cuarta– en su ciclo para la Universidad de Buenos Aires y participó en una reunión pública organizada por la comunidad judío-argentina para festejar la muy reciente inauguración de la Universidad Hebrea de Jerusalén y dinamizar la recolección de fondos para esa institución. A su muerte, Einstein legó a esta universidad los originales de sus documentos escritos, depositados por largos años en el Archivo de la Universidad de Princeton.

---

[10] A. Einstein Archive, Princeton University, Doc. 29-132, p. 20.
[11] A. Einstein Archive, Princeton University, Doc. 29-132, p. 20.

La visita a la clínica universitaria en compañía de José Arce, rector de la Universidad de Buenos Aires, quedó para el día siguiente; Arce y el establecimiento le dejaron a Einstein una excelente impresión. El miércoles 8 de abril, decidió adelantar el receso de Semana Santa de ese año y volver, con los Wassermann, a la estancia de Llavallol. Fuera del ruido de la ciudad, encontró allí "hermoso clima, maravillosa quietud" y escribió que había tenido "una idea sobre una nueva teoría sobre la conexión entre la gravitación y el electromagnetismo"[12].

El sábado 11 de abril la actividad continuó. Junto con varias personalidades locales, Einstein abordó vagones especiales del tren nocturno a Córdoba. Lo acompañaban el ya mencionado Isnardi, el ingeniero Enrique Butty, el físico Ramón Loyarte, el decano de la Facultad de Ingeniería, Luis A. Huergo, y el de la Facultad de Filosofía y Letras, Alberini. La prensa local anunció por anticipado que "[Einstein] será recibido [en Córdoba] como embajador espiritual de la nueva Alemania"[13] (haciendo referencia a la República de Weimar). Inmediatamente después de su llegada a Córdoba, las autoridades universitarias y provinciales lo invitaron a dar un paseo por las sierras. Visitó luego el lago San Roque, almorzó en el tradicional Edén Hotel de La Falda y regresó por el camino de Alta Gracia. De nuevo en la ciudad, pudo admirar la Catedral y los restos de lo que en su diario calificó como una antigua cultura. El lunes 13 de abril disertó durante media hora acerca del desarrollo de la teoría de la relatividad: la teoría restringida, la teoría general y los esfuerzos que contemporáneamente se hacían por poner la gravitación y el electromagnetismo dentro de un mismo esquema teórico. Un viejo amigo de Einstein, el doctor Georg Friedrich Nicolai, también pacifista, y que hasta tres años antes había sido profesor de fisiología en la Universidad de Berlín, enseñaba ahora en la Córdoba post-Reformista. Aunque ambos amigos se encontraron, es poco lo que se puede decir de esta entrevista. Una fotografía de grupo los muestra a ambos, pero a una cierta distancia uno del otro. Curiosamente el diario de Einstein no hace referencia alguna a ese encuentro y esto hace surgir la pregunta acerca de si el diario era realmente un reflejo de sus pensamientos "íntimos".

A su llegada, Einstein había expresado interés por visitar las colonias judías de Entre Ríos, pero ese deseo no pudo ser cumplido. Para su regreso de Córdoba a Buenos Aires, entonces, eligió hacer un viaje diurno (partió a las 6:45 de la mañana del día martes 14), para poder ver, por lo menos a través de la ventanilla del tren, el sur de esa provincia y parte de Santa Fe.

---

[12] A. Einstein Archive, Princeton University, Doc. 29-132, p. 21.
[13] "El profesor Alberto Einstein. Hoy llegará a Córdoba", *La Voz*, Córdoba, 12.04.1925.

Nuevamente en Buenos Aires, en la mañana del 16 de abril Einstein se reunió con dirigentes de la Federación Sionista Argentina. A la tarde asistió a una sesión especial de la Academia Nacional de Ciencias Exactas, Físicas y Naturales en su honor, en la que el presidente, el naturalista y escritor doctor Eduardo L. Holmberg (Ortiz, 1984, 2005; Bruno, 2011), le entregó su diploma de Académico Honorario. De acuerdo con la carta de invitación de Holmberg, a continuación se pasó a escuchar las preguntas que "los señores académicos y otras personas de conocida versación en la teoría de la Relatividad, pudieran hacer, al nuevo académico honorario, … relativas o ligadas con aquella teoría, rogando al doctor Einstein tuviese la deferencia de atenderlas"[14]. Los *Anales* de la Academia y otros textos históricos coinciden en destacar la intervención del joven físico de origen uruguayo Enrique Loedel Palumbo, luego maestro de importantes científicos. Su pregunta, relacionada con la existencia de una solución para un sistema de ecuaciones que describe el campo gravitacional de una masa puntual dio origen a una publicación en la revista alemana *Physikalische Zeitschrift* el año siguiente. Einstein encontró de interés esta pregunta; en cambio, sus impresiones, expresadas en su diario, sobre el resto de las intervenciones no dejan demasiado bien parados a sus interlocutores.

El 17 de abril, por la tarde, Einstein dictó su penúltima conferencia en la Universidad de Buenos Aires. Esa noche fue agasajado por el embajador alemán, en una reunión que comentó en su diario señalando que "sólo había locales, no alemanes".[15] La reunión tuvo lugar en un período en el que las tensiones políticas dentro de la comunidad alemana en la Argentina, y desde luego en Alemania, eran particularmente intensas. Entre los invitados argentinos se encontraban Ingenieros, el músico Carlos López Buchardo, el escritor Calixto Oyuela, el escultor Rogelio Yrurtia, el ingeniero Nicolás Besio Moreno, otras personalidades argentinas y, del lado argentino-alemán, miembros de la embajada, Wassermann y Ricardo Seeber, entonces presidente de la Asociación Cultural Argentino-Germana. Se debe tener en cuenta que Einstein había renunciado a su ciudadanía alemana antes de cumplir los 16 años, la edad de registro militar, y que más tarde, en 1901, había adquirido la nacionalidad suiza. Al ser incorporado a la Academia Prusiana de Ciencias implícitamente se le había reconocido nacionalidad alemana, aunque él nunca abandonó la adoptada. Al desatarse la Primera Guerra Mundial, 93 importantes intelectuales y científicos alemanes –entre otros, el microbiólogo Erlich, el químico Haber, el matemático Klein, los físicos Nernst, Planck y Röntegen– firmaron un manifiesto que daba apoyo firme y justificación a la guerra. Einstein, que era entonces profesor en la Universidad de Berlín, firmó un contra-manifiesto que denunciaba la guerra y la violencia en general como medio de resolver diferencias. Este documento solo tuvo el apoyo de otros tres nombres: Nicolai, el profesor recluido en Córdoba, posiblemente el autor de la declaración; Otto Bueck, que luego de la guerra representó a *La Nación* en Berlín; y un conocido astrónomo. A pesar de haber enfrentado numerosas dificultades, Einstein nunca dejó de reafirmar su fe pacifista; en su diario íntimo volcó su disgusto, comentando: "gente rara estos alemanes. Para ellos soy una flor maloliente, y sin embargo, una y otra vez, me ponen en su ojal" (cit. en Ortiz, 1995: 114).

---

[14] "Recepción del doctor Alberto Einstein en la sesión especial de la Academia del día 16 de abril de 1925", *Anales de la Academia Nacional de Ciencias Exactas Físicas y Naturales*, I. 1928, p. 322.
[15] A. Einstein Archive, Princeton University, Doc. 29-132, p. 33.

El sábado 18, por la tarde, Einstein ofreció una conferencia privada sobre su teoría en casa de sus anfitriones. Hacia la noche asistió, como invitado de honor, a una recepción en el cine-teatro Capitol donde expuso "Algunas reflexiones sobre la situación de los judíos". A continuación, la Asociación Hebraica le ofreció una recepción en sus salones y entregó a Einstein el diploma de socio honorario. El domingo, viajó nuevamente a Llavallol; por la noche se reunió con amigos alemanes residentes en Buenos Aires y luego asistió a una recepción en su honor ofrecida por dirigentes judío-argentinos en el Savoy Hotel.

Aún le quedaba por dar una conferencia el lunes 20 de abril, día en que anotó en su diario que había cumplido con su "última sesión científica con una audiencia entusiasta" (cit. en Ortiz, 1995: 115). También se hizo tiempo ese día para visitar, en su casa, al ingeniero Duclout que se encontraba enfermo. El martes 21, visitó el Hospital Israelita y otras organizaciones de caridad de la colectividad judía. Al mediodía, el Consejo Directivo de la Facultad de Ciencias Exactas lo invitó a almorzar en el Yacht Club Argentino, en el Tigre.

Entre todas estas intensas ocupaciones, Einstein logró también hacerse tiempo para componer algunos poemas breves que insertó, como dedicatoria, en fotografías suyas que luego obsequió a algunas de las personas con las que había hecho relación durante su viaje, como la señora de Wassermann, la escritora Elsa Jerusalem –esposa del profesor de embriología Víctor Widakowich, contratado en Viena por la Universidad de La Plata, a quien conoció en el *Cap Polonio* y con quien siguió manteniendo contacto– y el profesor Nirenstein. En sus dedicatorias, en alguna medida, Einstein definía los roles que cada uno de ellos jugó en su visita y les expresaba su simpatía y agradecimiento. La dedicatoria a Nirenstein sugiere que este desempeñó un papel importante en "moderar" a Einstein durante su estadía en la Argentina, "para que posiblemente nadie se sintiera ofendido" (cit. en Ortiz, 1995: 115).

El miércoles 22 fue invitado a un almuerzo de despedida organizado por la cúpula científica y política de Argentina en el Jockey Club, donde participaron rectores, decanos y ministros. El cierre de la jornada fue menos formal y algo más divertido: por la noche asistió a una fiesta organizada por el Centro de Estudiantes de Ingeniería en salones de la Asociación Cristiana de Jóvenes, que el homenajeado describió como "estudiantes, guitarras y canto". Entusiasmado por la cordialidad de la audiencia, Einstein aceptó tocar el violín. Según la crónica, un estudiante "con la velocidad de la luz" se disparó entonces a su casa en busca del instrumento.

Sería esperable que al día siguiente Einstein se hubiese levantado más tarde que de costumbre, aunque, quizás por modestia, este dato no consta en su diario. Al mediodía, los Wassermann ofrecieron en su casa un almuerzo de despedida, al que invitaron a los físicos Loyarte e Isnardi. El resto de ese día –el de su partida– el célebre científico lo ocupó en empacar su reducido equipaje (aumentado con algunos regalos) y en recibir a algunas amistades, los "íntimos", a los que obsequió las fotografías mencionadas. En la noche del 23 de abril, Einstein dejó para siempre la ciudad de Buenos Aires y se embarcó hacia Montevideo, adonde llegó –según reportan las crónicas de la época– con aspecto de cansado y sintiéndose algo enfermo.

En la capital uruguaya, Einstein ofreció tres conferencias en la Facultad de Ingeniería y fue invitado de honor en varias recepciones locales (Ortiz y Otero, 2001). Allí pudo gozar de una libertad que parece no haber tenido en Buenos Aires, donde es posible que haya estado más estrictamente controlado luego de algunas expresiones vertidas al comienzo de su visita. Por esta razón, la impresión que su diario refleja de los uruguayos es algo mejor que la que le dejaron los porteños. Luego de una semana, Einstein partió rumbo a Brasil en un buque "muy sucio y pequeño, pero con una tripulación agradable"[16]. Siempre "en el trapecio"[17], como se describiría Einstein a sí mismo en su diario, hizo una última parada breve en Rio de Janeiro. Finalmente, el 12 de mayo de 1925 emprendió su viaje de regreso a Europa.

Apenas llegado a Alemania, fue entrevistado por su amigo Otto Buek en su carácter de representante de *La Nación* en Berlín. En la entrevista publicada en ese diario el 5 de junio de 1925, Einstein pronosticó "un gran porvenir económico y cultural" para la Argentina y manifestó asimismo que conservaba "los mejores recuerdos de su hermoso viaje a la América del Sur" (*La Nación*, 5/6/1925). Su diario íntimo, sin embargo, no fue siempre fiel a sus declaraciones públicas y se cierra con una frase elocuente: "Al fin libre, pero más muerto que vivo"[18].

**5. Repercusiones de la visita**

El público local pudo percibir que el universo de ideas al que pertenecía Einstein anticipaba una importante revolución intelectual y es por eso que, aun sabiendo que la teoría de la relatividad emergía de hechos apenas perceptibles y que no parecía tener relación alguna con la vida cotidiana, escuchó al visitante con atención, respeto y admiración. Las ideas del físico no necesariamente fueron entendidas con fidelidad o en la misma forma que en Europa. Los reporteros de los diarios, justificadamente, no pudieron ir más allá de manifestar sus propias limitaciones, mientras que los filósofos intentaron expresar pensamientos coherentes, no siempre con éxito. Sin embargo, ambos grupos percibieron que esas ideas parecían indicar la necesidad de volver a reflexionar sobre la naturaleza del conocimiento científico y de la realidad física. En cuanto a los científicos de la Argentina, solo unos pocos podían seguir el complejo formulismo y las ideas que Einstein usaba en su teoría o en sus conferencias. Sin embargo, también los círculos académicos del país comprendieron que asistían a un acontecimiento poco común y mostraron la voluntad de hacer esfuerzos por no quedarse atrás y comprender.

---

[16] A. Einstein Archive, Princeton University, Doc. 29-132, p. 33.
[17] A. Einstein Archive, Princeton University, Doc. 29-132, p. 34.
[18] A. Einstein Archive, Princeton University, Doc. 29-132, p. 43.

La visita tuvo también otras consecuencias. Una de ellas, dentro de la comunidad universitaria y en los círculos de decisión sobre políticas culturales, fue la afirmación de la idea de que la investigación en las *ciencias teóricas* era un elemento cultural esencial que, injustamente, había sido desatendido en el pasado. La visita de Einstein tuvo lugar en un momento en el que se operaban cambios conceptuales importantes en las élites intelectuales de la Argentina, que las alejaban de los cánones de la filosofía positivista (o de una versión vernácula de esa filosofía) que habían sido dominantes por más de 20 años (Ortiz, 2011b; Gangui y Ortiz, 2011). En el campo de la ciencia, tanto en este país como en Europa, esa particular orientación filosófica había contribuido a impulsar la creación de grandes centros de investigación experimental. Así lo demuestran, en el caso específico de las ciencias exactas, el Instituto de Física de La Plata y el renovado y ambicioso Observatorio Astronómico de Córdoba.

Por otro lado, hay indicios de que esas preocupaciones existían ya antes de la llegada de Einstein y, de hecho, esto contribuye a explicar los esfuerzos por promover su visita. Efectivamente, en Buenos Aires ese cambio, paradigmático si se quiere, se asocia también, por ejemplo, con las conferencias que Ortega y Gasset dictó en la Facultad de Filosofía y con el sostenido apoyo que entonces comenzó a darse al desarrollo de las carreras de Ciencias Matemáticas y de Filosofía. Sin embargo, como se ha señalado, la insistencia de Alberini por lograr una definición de Einstein frente al positivismo y las ciencias físicas tropezó con el hecho de que una buena parte de los físicos mejor entrenados de la Argentina, como Loyarte e Isnardi, seguían mostrando una fuerte inclinación hacia lo que ellos entendían como positivismo (situación no muy diferente fuera de la Argentina).

La contratación del matemático puro Julio Rey Pastor luego de su exitosa visita de 1917 es un aspecto claro de la existencia de ese interés nuevo, o renovado, por las ciencias teóricas, que se prolongó a lo largo de casi toda la década de 1920. Además, poco después de la visita de Einstein se creó un departamento dedicado exclusivamente a las ciencias físicas en la Facultad de Ingeniería de Buenos Aires, donde los estudios teóricos recibieron una atención muy particular. La matemática moderna y luego la física-matemática comenzaron a desarrollarse en aquella Facultad como áreas más claramente independientes de investigación y se las consideró entonces no solamente por su utilidad indudable para la ingeniería sino también como capaces de contribuir a modificar la comprensión de la naturaleza e, incluso, a desafiar presupuestos filosóficos (Ortiz, 2012).

No puede decirse –y sería difícil esperarlo de una visita tan breve como la suya– que el viaje de Einstein determinó una transición clara y sostenida de la actividad científica argentina hacia la investigación en temas específicamente relacionados con la teoría de la relatividad. Esto último no había sido un objetivo claramente expresado, o siquiera tenido en cuenta, en el largo proceso que condujo al arribo de Einstein a las playas argentinas. Una política de envío de becados al exterior en el área de la teoría de la relatividad no había sido parte de la discusión, ni antes ni inmediatamente después de la visita. No debe olvidarse, por otro lado, que Loedel Palumbo había sido introducido a las ideas de la relatividad por su maestro Gans, en La Plata, *antes* de esa visita. Si bien aquel joven físico tuvo luego discípulos, hacia fines de la década de 1920 encontró resistencias; sus propios intereses se trasladaron hacia el estudio de los fundamentos de las ciencias físicas. Además, en la década de 1930 el interés por la nueva mecánica cuántica restó impulso, a nivel internacional, al esfuerzo en favor de la relatividad. Sin embargo, no cabe duda de que la presencia del célebre científico fue un serio espaldarazo para la física en la Argentina y una fuerte motivación para los jóvenes en formación, especialmente para aquellos que pudieron presenciar los esfuerzos del visitante para comunicarse y compartir con ellos sus visiones del futuro de la ciencia.

Esa visita tuvo también un impacto positivo sobre el temario de los cursos universitarios más avanzados de ciencias exactas. El caso de Rey Pastor es paradigmático: incluyó temas propios de la matemática de la relatividad, como la geometría diferencial moderna, en sus seminarios de matemática avanzada en 1923 y 1924, poco antes de la visita (Rey Pastor, 1989). Más adelante, en la década de 1930, cuando se estaba forjando una importante generación de físicos teóricos en la Argentina (Ortiz, 2012), volvió a incluir tópicos relativistas en sus seminarios avanzados, como el álgebra tensorial y la relatividad especial y, luego, el cálculo tensorial y la relatividad general en 1936. Por su parte, Butty, también en la década de 1930, introdujo en su curso de física-matemática en la Universidad de Buenos Aires el estudio sistemático del cálculo tensorial, una de las herramientas básicas de la teoría de la relatividad (Butty, 1931, 1934).

Por todas estas razones puede decirse, sin duda, que en la lista de los viajeros científicos que llegaron a la Argentina en el siglo XX la visita de Einstein ocupa una posición de considerable interés e importancia, tanto a causa de su impacto sobre la matemática, la física y la filosofía, como por el que tuvo en áreas muy diversas de la cultura, independientemente de que ese impacto haya sido o no fiel a la teoría de la relatividad.

**Fuentes**

Butty, E. (1931, 1934). *Introducción a la física matemática I, y II*, Buenos Aires, Imprenta de la Universidad.

Duclout, J. (1920), "Materia, energía, relatividad", *Revista del Centro de Estudiantes de Ingeniería*, n° 21, pp. 628-643.

Einstein, A. (2008), "Unpublished opening lecture for the course on the theory of relativity in Argentina, 1925" (trad. de Gangui, A. y Ortiz, E. L.), *Science in Context*, vol. 21, n° 3: pp. 451-459.

Lugones, L. (1921), *El tamaño del espacio; ensayo de psicología matemática*, Buenos Aires, El Ateneo.

Meyer, C. (1915), "La radiación y la teoría de los quanta", *Anales de la Sociedad Científica Argentina*, vol. 8, pp. 5-103, 153-245 y 281-371.

Rey Pastor, J. (1989), *The Works of Julio Rey Pastor* [ed. E. L. Ortiz], vol. I-VIII, Londres, The Humboldt Press.

Volterra, V. (1910), "Espacio, tiempo i masa", *Anales de la Sociedad Científica Argentina*, n° 70, pp. 223-243.

**Bibliografía**

Asúa, M. de y Hurtado de Mendoza, D. (2006), *Imágenes de Einstein*. Buenos Aires, Eudeba.

Bruno, P. (2011), *Pioneros culturales de la Argentina. Biografías de una época, 1860-1910*, Buenos Aires, Siglo Veintiuno Editores.

Galles, C. (1982), "La repercusión en la Argentina de las teorías relativistas (1905-1925)", *Actas de las Primeras Jornadas de Historia del Pensamiento Científico Argentino*, Buenos Aires, Ediciones FEPAI.

Gangui, A. (comp.) (2007), *El universo de Einstein*, Buenos Aires, Eudeba.

Gangui, A. y Ortiz, E. L. (2005), "Crónica de un mes agitado: Albert Einstein visita la Argentina", *Todo es Historia*, n° 454, pp. 22-30.

— (2008), "Einstein's Unpublished Opening lecture for his Course on Relativity Theory in Argentina, 1925", *Science in Context*, vol. 21, n° 3: 435-450.

— (2009), "First echoes of Relativity in Argentine astronomy", en Romero, G., Cellone, S. y Cora S. (eds.), *Historia de la Astronomía Argentina*, La Plata, AAABS n° 2 (suplemento), pp. 31-37.

— (2011), "Anti-positivismo, ciencias teóricas y relatividad en la Argentina de la década de 1920", *Revista Brasileira de História da Ciência*, vol. 4, n° 2, pp. 201-218.

Ortiz, E. L. (1984), "La polémica del Darwinismo y la inserción de la ciencia moderna en la Argentina. Conferencia de Clausura del III Congreso de la Sociedad Española de Historia de la Ciencias", *Actas*, vol. I, pp. 89-108.

— (1988), "Las relaciones científicas entre Argentina y España a principios de este siglo: La Junta para Ampliación de Estudios y la Institución Cultural Española", en Sánchez Ron, J. M. (ed.), *La Junta para Ampliación de Estudios e Investigaciones Científicas 80 años después*, Madrid, Consejo Superior de Investigaciones Científicas, pp. 119-158.

— (1995), "A convergence of interests: Einstein's visit to Argentina in 1925", *Ibero-Americanisches Archiv*, n° 21, pp. 67-126.

— (2005), "On the Transition from Realism to the Fantastic in Argentine Literature of the 1870s: Holmberg and the Córdoba Six", en Fishburn, E y Ortiz, E. L. (eds.), *Science and the Creative Imagination in Latin America*, Londres, Institute for Advanced Study, pp. 59-85.

— (2011a), 'Julio Rey Pastor, su posición en la escuela matemática argentina'. *Revista de la Unión Matemática Argentina*, 52 (1): 149-194.

— (2011b), 'The emergence of theoretical physics in Argentina: Mathematics, mathematical physics and theoretical physics, 1900-1950'. En L. Brink y V. Mukhanov, (eds.). Héctor Rubinstein Memorial, Trieste: SISSA, pp. 13-34.

— (2012), "Julio Rey Pastor y los físicos. Matemática, física-matemática, física teórica: 1925-1935", en Hurtado, D. (comp.), *La Física y los físicos argentinos. Trayectorias, espacios institucionales y memorias*, Córdoba, Universidad Nacional de Córdoba, pp. 397-441.

Ortiz, E. L. y Otero, M. H. (2001), "Removiendo el ambiente: La visita de Einstein al Uruguay en 1925", *Mathesis*, serie II, n° 1, pp. 1-35.

Ortiz, E. L. y Rubinstein, H. (2009), "La física en la Argentina en los dos primeros tercios del siglo veinte: algunos condicionantes exteriores a su desarrollo", *Revista Brasileira de História da Ciência*, vol. 2, n° 1, pp. 40-95.

Pyenson, L. (1985), *Cultural Imperialism and the Exact Sciences*, New York, Peter Lang.

Seelig, C. (1954), *Albert Einstein; eine dokumentarische Biographie*, Zürich, Europa Verlag.